\documentclass[aps,prd,twocolumn,letterpaper,showpacs,floatfix]{revtex4}
\usepackage{epsfig}
\usepackage{graphicx}
\usepackage{amsmath}
\usepackage{color}
\usepackage{psfrag}
\newcommand{\pderiv}[2]{\frac{\partial #1}{\partial #2}}
\newcommand{\create}[1]{a_{\vec{ #1}}^\dagger}
\newcommand{\uncreate}[1]{a_{\vec{ #1}}}
\newcommand{\npm}{n^{\pm}}

\newcommand{\mpm}{m^{\pm}}
\newcommand{\bra}[1]{\langle #1|}
\newcommand{\ket}[1]{|#1\rangle}
\newcommand{\braket}[2]{\langle #1|#2\rangle}
\newcommand{\suminf}[1]{\sum_{#1 = 0}^\infty}

\newcommand{\PlMass}{M_{\rm Pl}}
\newcommand{\trpsq}{1-\rm{Tr}(\rho^2)}
\newcommand{\mode}[2]{g^{#2}(#1,\tau)}
\newcommand{\modedot}[2]{\pderiv{\mode{#1}{#2}}{\tau}}
\renewcommand{\vec}[1]{ {\bf #1}}
\def\rmmat#1{{\hbox{\rm #1}}}
\def\rmscr#1{\rmmat{\scriptsize #1}}
\def\ie{{\em i.e.}\ }

\begin{document}
\title{Characterizing Entanglement Entropy Produced by Non-Linear
  Scalar Interactions During Inflation} \author{Dan Mazur and Jeremy
  S. Heyl\footnote{Canada Research Chair}} \affiliation{Department of
  Physics and Astronomy,
  University of British Columbia \\
  6224 Agricultural Road, Vancouver, British Columbia, Canada, V6T
  1Z1} \date{\today}
\begin{abstract}

  The density fluctuations that we observe in the universe today are
  thought to originate from quantum fluctuations produced during a
  phase of the early universe called inflation.  By evolving a
  wavefunction describing two coupled Fourier modes of a scalar field
  forward through an inflationary epoch, we demonstrate that
  non-linear effects can result in a generation of entanglement
  entropy between modes with different momenta in a scalar field
  during the inflationary period when just one of the modes is
  observed.  Through this mechanism, the field would experience
  decoherence and appear more like a classical distribution today;
  however the mechanism is not sufficiently efficient to explain
  classicality.  We find that the amount of entanglement entropy
  generated scales roughly as a power law $S \propto \lambda^{1.75}$,
  where $\lambda$ is the coupling coefficient of the non-linear
  potential term.  We also investigate how the entanglement entropy
  scales with the duration of inflation and compare various
  entanglement measures from the literature with the von Neumann
  entropy.  This demonstration explicitly follows particle creation
  and interactions between modes; consequently, the mechanism
  contributing to the generation of the von Neumann entropy can be
  easily seen.
\end{abstract}
\maketitle

\section{Introduction}

Most modern cosmological models include a period in the universe's
history called inflation during which the scale parameter increased
exponentially with the proper time of a comoving observer.  This
period was originally introduced to address the horizon and flatness
problems of cosmology \cite{1981PhRvD..23..347G}.  More recently,
however, research on inflation has been toward understanding structure
formation
\cite{1982PhRvL..49.1110G,1982PhLB..117..175S,1985PhRvD..31.1792L}.
The distribution of galaxies and clusters that we observe in the
universe today are thought to have originated from fluctuations of a
quantized field created during inflation~\cite{1981JETPL..33..532M,Hawking:1982}.
A thorough review of structure formation and inflationary cosmology
can be found in Liddle and Lyth~\cite{Liddle:2000}.

Despite their quantum mechanical origins, the late-time evolution of
these fluctuations is treated in a classical framework.  It is
therefore important to understand the quantum-to-classical transition
made by these fluctuations (for a recent review, see
~\cite{Kiefer:2008ku}).  The classicality of a quantum system is often
discussed in the context of decoherence.  That is, as a quantum system
interacts with unobserved environmental influence, that system loses
quantum coherence and begins to behave as a classical statistical
distribution.

The quantized field may of course be the inflaton itself, which drives
the inflation of the universe, or it could be another quantized field
that produces density fluctuations as in curvaton models or the
gravitational field.  It is possible, in principle, that
non-classical correlations from an inflationary period in our
universe's history may one day be observed.  But this depends on the
decoherence that the scalar or tensor field has experienced since the
beginning of inflation. Several authors have investigated decoherence
of the density fluctuations by calculating the entropy of cosmological
perturbations created during inflation
~\cite{2000PhRvD..62d3518K,2005PhRvD..72d5015C,2006PhRvD..74b5001C,
  2007JCAP...11..029P,campo:065044,2008PhRvD..78f5045C}.

It has been suggested~\cite{Burgess:2006} that decoherence is unlikely
to occur during inflation because the Bunch-Davies state occupied by
the scalar field during inflation is similar to the Minkowski vacuum.
Because the ordinary Minkowski vacuum does not decohere, we would not
expect to see any decoherence from a scalar field during inflation.
In the particle-based picture adopted for the present analysis, it
becomes clear that the scalar field does undergo decoherence when
the potential is non-linear.  

Since decoherence is a necessary condition for the emergence of
classicality in a quantum system \cite{Zurek:1993}, non-linearities in
the scalar field help to explain the classical matter distribution
that we observe today.  This simple model demonstrates that this
entropy generation can occur during inflation itself and does not
depend on the reheating process at the end of inflation; therefore,
the results are perhaps most interesting for cosmological scalar
fields that do not participate in reheating.  For such fields, the
non-linear interactions do not generate a sufficient amount of
decoherence to result in classicality for the fields.

Here, we examine the case where certain modes of a field play the role
of the environmental influence and cause decoherence when a
non-linearity in the potential allows the modes to
interact~\cite{Lombardo:1995,Burgess:2006,Martineau:2006}.  We discuss
a simulation that was performed to compute the entanglement entropy
between such modes in a very transparent model that follows particle
creation and the interaction between modes during the inflationary
period.  The entropy is computed as inflation progresses to
demonstrate the decoherence of a scalar field.  

Computing the entanglement entropy of a large quantum system is a
computationally difficult task since it involves diagonalizing the
density matrix.  To evaluate several possible expediencies, we have
compared our results to other measures of entanglement and
correlations between modes.  We have found that the other measures
considered share a similar qualitative behaviour with the
entanglement entropy and can be much easier to compute.  Therefore,
for some applications, these measures may be useful as stand-in
quantities in simulations where the entanglement entropy is too costly
to compute.  We verify several efficient methods to characterize the
entropy.

\section{Cosmological Scalar-Field Evolution}

We would like to investigate the evolution of a scalar field in an
isotropic, homogeneous, flat spacetime.  The analysis for this
situation is covered extensively in part I, chapter 6 of Mukhanov et
al. \cite{Mukhanov:1990me}.  The relevant metric for this evolution is
\begin{equation}
	\label{eq:inflFRW}
	ds^2 = a^2(\tau)(d\tau ^2 - d{\bf x}^2).
\end{equation} 
where $\tau$ is the conformal time, which is related to the comoving time 
by $dt=a(\tau) d\tau$, and ${\bf x}$ is a comoving displacement.  For
simplicity we will take $a(\tau)=-(H\tau)^{-1}$ (pure deSitter
expansion) during inflation.

The evolution of a scalar field $\phi$ is governed by its Lagrangian
$\mathcal{L}$.  The lowest-order Lorentz-invariant expression
containing up to first derivatives is
\begin{equation}
	\label{eq:inflsfl}
	\mathcal{L} = \frac{1}{2} g^{\mu \nu} \partial_\mu \phi \partial_\nu\phi - V(\phi).
\end{equation}
For simplicity we will neglect the mass of the scalar field
during inflation ($m \ll H$).  We include a non-linearity in the
potential that couples the Fourier modes of the field.  Even if the
field itself is free, its self-gravity will introduce an interaction
potential of the form ~\cite{Burgess:2006, Martineau:2006}
\begin{equation}
	\label{eq:inflpotential}
	V=\lambda \PlMass \phi^3.
\end{equation}
Although the $\phi^3$ potential is generally unstable, one should
interpret this as an effective potential to account for the
gravitational self-interaction, so the instability is not surprising
because the gravitational self-interaction is generally unstable.

\subsection{Mode coupling during inflation}

For this analysis, we choose to use a simple model in which the universe
contains only particles with four possible momenta: $\pm \vec{k}$ and
$\pm 2\vec{k}$.  Given this requirement, we construct a Hamiltonian which
incorporates a coupling term between these two Fourier modes so that
we can observe the effect this non-linearity has on the entanglement 
between modes during inflation.

The creation and annihilation operators satisfy the following
commutation relations
\begin{equation}
	\label{eq:inflcommute}
	[\uncreate{k},\create{k'}]=\delta^{(3)}(\vec{k}-\vec{k'})
\end{equation}
\begin{equation}
	\label{eq:inflcommute2}
	[\create{k},\create{k'}] = [\uncreate{k},\uncreate{k'}]=0.
\end{equation}

Including our potential term (\ref{eq:inflpotential}), the action for
the field is
\begin{equation}
   S = \frac{1}{2}\int d^4x \sqrt{-g}\biggr [\partial_{\mu} \phi
   \partial^{\mu}\phi 
        + \lambda \PlMass \phi^3\biggr].
\end{equation}

Following the steps outlined by ref.~\cite{Heyl:2006fb}, we arrive at
the following expression for the action.
\begin{equation}
	S = \frac{1}{2}\int d^4x a^2\left\{ \left[\pderiv{\phi}{\tau}\right]^2
	- [\nabla \phi^2] 
	+ a^2\lambda \PlMass  \phi^3\right\}
\end{equation}

\begin{widetext}
If we make the substitution $u=a\phi=\frac{1}{(2 \pi)^{3/2}}
\int d^3k u_{k}(\tau)e^{i \vec{k}\cdot \vec{r}}$, the action becomes
\begin{equation}
	S = \frac{1}{2}\int d\tau d^3k \left[ \left|\pderiv{u_k}{\tau}\right|^2
	- (k^2+m_{\rm eff}^2)|u_k|^2 - \frac{ \lambda \PlMass}{(2 \pi)^{1/2}a} \int{d^3k' d^3k''
	u_{\vec{k}} u_{\vec{k}'} u_{\vec{k}''}\delta^{(3)}(\vec{k} + \vec{k}' + \vec{k}'')}
	\right]
\end{equation}
where the effective mass is $m_{\rm eff}^2 = -2\frac{Q}{\tau^2}$, 
\begin{equation}
	Q \equiv \frac{1}{(1+3w)^2}\left[(1-3w)\right]
\end{equation}
and $w=p/\rho$ is the equation of state parameter.

The Hamiltonian is, then,
\begin{equation}
	\label{eq:Ham1}
	H = \frac{1}{2}\int d^3k\left[\left|\pderiv{u_{\vec{k}}}{\tau}\right|^2
	+ (k^2+m_{\rm eff}^2)|u_{\vec{k}}|^2 + \frac{ \lambda \PlMass}{\sqrt{2 \pi}a}
	\int d^3k' u_{\vec{k}}u_{\vec{k}'}u_{-(\vec{k}+\vec{k}')}\right].
\end{equation}
In general, we have $u_{\vec{k}} = g(k,\tau)\uncreate{k} + g^* (k,\tau)
\create{-k}$.  Putting this into to the Hamiltonian, (\ref{eq:Ham1}), and 
neglecting terms that do not conserve energy in flat spacetime gives
\begin{eqnarray}
	\label{eq:Ham2}
	H &=& \frac{1}{2}\int d^3k\biggr[\left(\left|\modedot{k}{}\right|^2 +|\mode{k}{}|^2(k^2+m_{\rm eff}^2)\right)
	(\create{k}\uncreate{k} + \uncreate{k}\create{k}) \nonumber \\
	&+& 
	\left(\modedot{k}{*2} +\mode{k}{*2}(k^2+m_{\rm eff}^2)\right)\create{-k}\create{k} +
	\left(\modedot{k}{2} +\mode{k}{2}(k^2+m_{\rm eff}^2)\right)\uncreate{k}\uncreate{-k}  \nonumber \\ 
	& + & \frac{ \lambda \PlMass}{\sqrt{2 \pi}a}
	\left[\mode{k}{}\mode{k}{}\mode{2k}{*}\create{2k}\uncreate{k}\uncreate{k} +
		\mode{k}{*}\mode{k}{*}\mode{2k}{}\uncreate{2k}\create{k}\create{k}\right]
	\biggr]
\end{eqnarray}

The mode function is normally chosen to be 
\begin{equation}
	\label{eq:usualmode}
	\mode{k}{} = -\frac{1}{\sqrt{2 k^3}}(i-k\tau) \frac{e^{-ik\tau}}{\tau} 
\end{equation}
as this choice satisfies the equation of motion for the free field
during a deSitter phase and because it simplifies the Hamiltonian to
one that commutes with the number operator since, when $Q=1$,
\begin{equation}
\modedot{k}{2}+\mode{k}{2}(k^2+m_{\rm eff}^2)=0.
\end{equation}
However, this choice is not practical for our calculation because the
scalar field is not free; therefore, this choice does not satisfy the
field equation of motion, and in fact it complicates the Hamiltonian
because, for example
\begin{equation}
	\mode{k}{}\mode{k}{}\mode{2k}{*} = \frac{1}{4 k^{9/2}\tau} (2(k\tau)^3-3i(k\tau)^2-i)
\end{equation}
does not have a simple dependence on $\tau$ and the simplifications provided by 
(\ref{eq:usualmode}) are lost.

We would like to know the amount of entropy at the end of inflation during radiation domination.  
The usual way to proceed is to select the mode function ($\ref{eq:usualmode}$) 
and use this to determine the equation of motion for the scalar field during 
inflation.  We would then determine the Bogoliubov 
coefficients at the transition from inflation to radiation domination.  
After performing the transformation, we would compute 
the amount of entropy from the transformed density matrix.

However, we can simplify the problem by instead choosing a mode function 
that describes the system during radiation domination and use this 
mode function to compute the entire evolution.
The choice of function $g(k,\tau)$ is flexible due to the vacuum
ambiguity and is related to choosing the set of states that the
creation and annihilation operators act upon.  Any choice will provide
us with a complete basis with which we can describe any state of the
field.  The arrbitrariness of the mode function is also discussed in~\cite{2007JCAP...02..031A}.
  
For us, it is most prudent to choose the simple function
\begin{equation}
\label{eq:modefunction}
g(k,\tau)=\frac{1}{\sqrt{2k}}e^{-ik\tau} 
\end{equation}
which defines the vacuum
both during radiation domination and for scales much smaller than the
horizon even during the de Sitter phase.  Thus, we can make a very natural 
connection between our initial state and our final state.  
The choice is as arbitrary as choosing
to perform a calculation in classical mechanics in a rotating frame
rather than an inertial frame.

Correctly interpreting the wavefunction where (\ref{eq:modefunction})
is inappropriate (i.e. after horizon exit during a de Sitter phase)
would require a Bogoliubov transformation, but for our purposes we do
not require this.  We are only interested in calculating the entropy
after the transition to radiation domination where our choice of mode
function corresponds to the usual creation and annihilation operators
for this background.  Therefore, we avoid transformations entirely
since we already have the required description of our wavefunction.

 With the choice (\ref{eq:modefunction}), the Hamiltonian is not constant in
time even without the non-linear couplings.  In particular the mass
depends on time; this choice is similar in spirit to the calculations
of Guth and Pi~(\cite{1982PhRvL..49.1110G}).  Heyl~\cite{Heyl:2006fb}
has shown that for a free scalar field that this choice gives the same
results as the standard function $g(k,\tau)$ and we refer the reader
to that article for a more thorough discussion of the technique.
 
Choosing to use ($\ref{eq:modefunction}$), we have
\begin{equation}
u_{\vec{k}} = \frac{1}{\sqrt{2 k}}(e^{-ik \tau}\uncreate{k} 
	+ e^{ik \tau}\create{-k}) .
\end{equation}
The nonlinear terms in the Hamiltonian provide a coupling mechanism
between the modes of interest.  To perform the integral over $d^3k'$
in (\ref{eq:Ham1}), we neglect the effect of the coupling on the modes
that are not considered in our simulation and treat the functions
$u_\vec{k}$ as constant on a spherical shell surrounding the momenta,
$\vec{k}'$, that we are interested in.  For
$u_{\vec{k}'}=\text{const.}$ on spherical shells of constant volume
around $\vec{k}$ and $2\vec{k}$, the integral becomes
\begin{equation}
\int d^3k' u_{\vec{k}}u_{\vec{k}'}u_{-(\vec{k}+\vec{k}')} \rightarrow Vk^3u_{\vec{k}}u_{\vec{k}}u_{-2\vec{k}}
\end{equation}
where $V=\frac{4}{3}\pi\left(\frac{4}{1+\sqrt[3]{2}}\right)^3
\approx23$ is a (somewhat arbitrary) geometrical constant.

Making this substitution, we arrive at the final form of the Hamiltonian.
\begin{equation}
	\label{eq:inflHam}
	H = \int d^3\vec{k}\biggr[\left(k-\frac{Q}{\tau ^2 k} \right)(\create{k}\uncreate{k}+
	\uncreate{k}\create{k}) -\frac{Q}{\tau ^2 k}
	( \uncreate{-k}\uncreate{k} e^{-2ik\tau} + \create{-k}\create{k} e^{2ik\tau})+ \frac{\lambda V k^{3/2} \PlMass}{4\sqrt{2\pi}a}(\create{2k}\uncreate{k}\uncreate{k} 
        + \uncreate{2k}\create{k}\create{k})\biggr].
\end{equation}
This Hamiltonian is similar to that used by ref.~\cite{Heyl:2006fb},
generalized to allow for the interactions between Fourier modes.

Here, the two terms multiplied by the factor $\lambda$ are responsible
for the annihilation of two particles from the $\vec{k}$ mode into a
single particle from the $2\vec{k}$ mode and the decay of an
$2\vec{k}$ mode particle into two $\vec{k}$ mode particles,
respectively. 
As the two modes of the field exchange particles with each
other, we expect that entanglement entropy will be generated in either
of the modes observed individually.

We wish to use this Hamiltonian to evolve Fock space wavefunctions
representing the number of particles in each of four modes: Those with
$m^+$ particles with momentum $2\vec{k}'$, $m^-$ particles with
momentum $-2 \vec{k}'$, $n^+$ particles with momentum $\vec{k}'$, and
$n^-$ particles with momentum $-\vec{k}'$.
\begin{eqnarray}
	\label{eq:inflwavefunctionlong}
	\ket{\psi} &=& \suminf{m^+} \suminf{m^-} \suminf{n^+} \suminf{n^-} B_{m^+, m^-, n^+, n^-}(\tau)
	\left(\frac{(\create{2k'})^{m^+}}{\sqrt{m^+}[\delta^{(3)}(2\vec{k}'-2\vec{k}')]^{\frac{m^+}{2}}}\right)
	\left(\frac{(\create{-2k'})^{m^-}}{\sqrt{m^-}[\delta^{(3)}(2\vec{k}'-2\vec{k}')]^{\frac{m^-}{2}}}\right)
	\nonumber \\
	& & \times
	\left(\frac{(\create{k'})^{n^+}}{\sqrt{n^+}[\delta^{(3)}(\vec{k}'-\vec{k}')]^{\frac{n^+}{2}}}\right)
	\left(\frac{(\create{-k'})^{n^-}}{\sqrt{n^-}[\delta^{(3)}(\vec{k}'-\vec{k}')]^{\frac{n^-}{2}}}\right)
	\ket{0} \\
	&=& \suminf{m^+, m^-, n^+, n^-}\!\!\!\!\!\!\!\!\!\!\!\!
	B_{m^+, m^-, n^+, n^-}(\tau)
	\ket{m^+,2\vec{k}'}\ket{m^-,-2\vec{k}'}\ket{n^+,\vec{k}'}\ket{n^-,\vec{-k}'}.
\end{eqnarray}
Whenever possible, we will use simplified notation such as
\begin{equation}
	\label{eq:inflWavefunc}
	\ket{\psi} = \sum_{\npm,\mpm=0}^\infty B_{\npm,\mpm}(\tau) \ket{\mpm}\ket{\npm}.
\end{equation}

In order to evolve the wavefunction forward in time, we replace $\tau$
with a new variable, $x=-1/(k\tau)$.  The equation of motion is then
found from $i\frac{d}{d\tau}\ket{\psi}= H \ket{\psi}$, left multiplied
by $\bra{\npm, \pm \vec{k}}\bra{\mpm , \pm 2\vec{k}}$. The following
identities are needed to evaluate $H\ket{\psi}$:
\begin{eqnarray}
	\label{eq:inflcreateidentity1}
	\create{k}\uncreate{k}\ket{\mpm}\ket{\npm} &=& [m^+\delta^{(3)}(\vec{k}-2\vec{k}') + n^+ \delta^{(3)}(\vec{k} -\vec{k}') 
	+m^- \delta^{(3)}(\vec{k}+2\vec{k}') +n^-\delta^{(3)}(\vec{k}+\vec{k}')]
	\ket{\mpm}\ket{\npm} \\
	\label{eq:inflcreateidentity2}
	(\create{k}\uncreate{k}+\uncreate{k}\create{k})\ket{\psi} &=& 
	(2\create{k}\uncreate{k}+Z)\ket{\psi} \\
	\label{eq:inflcreateidentity3}
	\uncreate{-k}\uncreate{k} \ket{\mpm}\ket{\npm}&=&\sqrt{m^+m^-}(\delta^{(3)}(\vec{k}-2\vec{k}') + \delta^{(3)}(\vec{k}+2\vec{k}'))\ket{\mpm-1}\ket{\npm}  \nonumber\\
	& &  + \sqrt{n^+n^-}(\delta^{(3)}(\vec{k}-\vec{k}') + \delta^{(3)}(\vec{k}+\vec{k}'))\ket{\mpm}\ket{\npm-1} \\
	\label{eq:inflcreateidentity4}
	\create{-k}\create{k} \ket{\mpm}\ket{\npm}&=&\sqrt{(n^++1)(n^-+1)}(\delta^{(3)} (\vec{k}-\vec{k}')+\delta^{(3)}(\vec{k}+\vec{k}'))\ket{\mpm}\ket{\npm+1}\nonumber\\
	& &  + \sqrt{(m^++1)(m^-+1)}(\delta^{(3)}(\vec{k}-2\vec{k}') +
	\delta^{(3)}(\vec{k}+2\vec{k}'))\ket{\mpm+1}\ket{\npm} \\
	\label{eq:inflcreateidentity5}
	\uncreate{2k}\create{k}\create{k} \ket{\mpm}\ket{\npm} &=&
	\sqrt{m^+(n^++1)(n^++2)}\delta^{(3)}(2\vec{k}-2\vec{k}')\ket{m^+-1}\ket{m^-}\ket{n^++2}\ket{n^-} \nonumber \\
	& &
	+\sqrt{m^-(n^-+1)(n^-+2)}\delta^{(3)}(2\vec{k}+2\vec{k}')\ket{m^+}\ket{m^--1}\ket{n^+}\ket{n^-+2} \\
	\label{eq:inflcreateidentity6}
	\create{2k}\uncreate{k}\uncreate{k} \ket{\mpm}\ket{\npm} &=&
	\sqrt{n^+(n^+-1)(m^++1)}\delta^{(3)}(\vec{k}-\vec{k}')\ket{m^++1}\ket{m^-}\ket{n^+-2}\ket{n^-} \nonumber \\
	& & +\sqrt{n^-(n^--1)(m^-+1)}\delta^{(3)}(\vec{k}+\vec{k}')\ket{m^+}\ket{m^-+1}\ket{n^+}\ket{n^--2}
\end{eqnarray}
where $Z=[\uncreate{k},\create{k}]=\delta^{(3)}(\vec{k}-\vec{k})$ is
an infinite constant.

After some algebra, we find the time evolution of the states is given by 
\begin{eqnarray}
	\label{eq:inflEOM}
	i \frac{d}{dx}A_{\mpm,\npm}(x) &=&
	-\frac{Q}{2}\biggr[\biggr(\sqrt{m^+ m^-}A_{\mpm-1,\npm} 
	+\sqrt{n^+n^-}A_{\mpm,\npm -1}\biggr)e^{-2i\gamma/x}\nonumber \\ 
	& &   +\biggr(\sqrt{(n^+ +1)(n^-+1)}A_{\mpm,\npm+1} 
	+\sqrt{(m^++1)(m^-+1)}A_{\mpm+1,\npm}\biggr)e^{2i\gamma/x}\biggr]
	\nonumber \\
	& &     +\frac{\alpha}{x^3}\biggr[
	\biggr(\sqrt{(n^+-1)(n^+)(m^++1)}A_{m^++1,m^-,n^+-2,n^-} \nonumber\\
	& & +   \sqrt{(n^--1)(n^-)(m^-+1)}A_{m^+,m^-+1,n^+,n^--2}\biggr) \nonumber\\
	& &     +
	\biggr(\sqrt{(m^+)(n^++1)(n^++2)}A_{m^+-1,m^-,n^++2,n^-}\nonumber\\  
	& & +\sqrt{(m^-)(n^-+1)(n^-+2)}A_{m^+,m^--1,n^+,n^-+2}\biggr)\biggr]
\end{eqnarray}
where the matrices $A$ and $B$ are related by a phase transformation
\begin{equation}
A_{\mpm,\npm}(x)= e^{-i (m^++m^-+n^++n^-+Z) (\gamma(x)-1)/x}
B_{\mpm,\npm}(x)
\end{equation}
with $\gamma(x)=2+Qx^2$ and $Z$ is an infinite constant (related to
the renormalization of the vacuum energy).  The dimensionless constant
$\alpha$ has the value $\frac{\lambda V H}{8 \sqrt{2\pi k \phi}}$.  To
arrive at equation (\ref{eq:inflEOM}), we have ignored terms that
involve modes $\pm \frac{1}{2}\vec{k}$ and $\pm 4\vec{k}$ since we are
not concerned with how these modes evolve for our present purposes.
\end{widetext}

We begin the simulation for small values of $x$, well before the modes
cross outside the Hubble length.  At such a time, there has been a
negligible amount particle production, so our initial wavefunction is
simply the Fock vacuum, $\ket{\psi}_i = \ket{\mpm=0}\ket{\npm=0}.$ In
the limit of $\frac{k}{a} \ll H$ or $x \ll 1$, this initial condition
corresponds to the Bunch-Davies vacuum.  During
vacuum-energy-domination, the equation of state parameter, $w$, equals $-1$.
Therefore, neglecting the mass of the scalar field, the value of $Q$
is unity.


\subsection{Entanglement measures}

Discussions of decoherence rely on the notion of an environment: a
collection of degrees of freedom that interacts and becomes entangled
with the system of interest.  Our model is naturally separated into
modes with different magnitudes of momentum.  Noting that the
entanglement entropy does not depend on our choice of which set of
modes is the environment and which is the system, we identify the
modes with momentum $\pm 2\vec{k}$ with the environmental degrees of
freedom and the modes with momentum $\pm \vec{k}$ to be the system.

This choice represents an entanglement due to coarse graining the
internal degrees of freedom of the scalar field based on scale.  One
can think of the coarse graining as either being due to practical
limitations in the observations that can be made or as physical
limitations such as a mode being entangled with a mode with a
wavelength greater than the horizon size.  The latter case is
discussed in ~\cite{Martineau:2006}.

\begin{figure}[h]
	\centering
		\includegraphics[width=8.5cm]{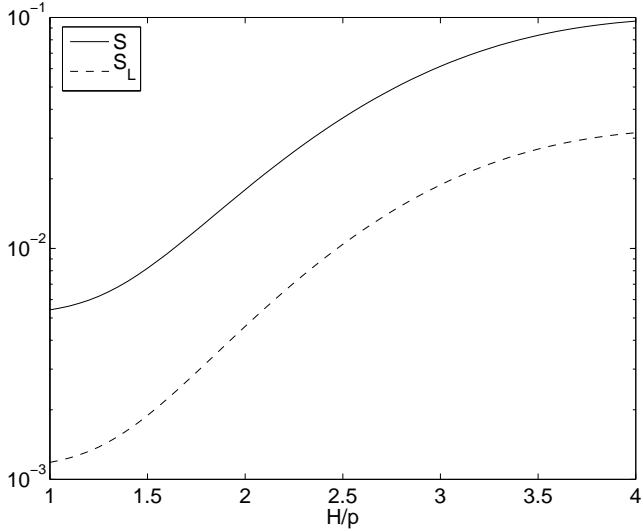}
                \caption{The evolution of entanglement entropy, $S$,
                  and linear entropy, $S_L=|1-{\rm Tr}(\rho^2)|$, as
                  the mode stretches past the horizon for
                  $\alpha=0.2$.  This demonstrates that the
                  non-linearities in the inflaton potential are
                  capable of producing entanglement entropy between
                  the coupled modes.  Also note that $S$ scales
                  monotonically with $S_L$.}
	\label{fig:Sandtrrho2}
\end{figure}

We measure the entanglement between modes using two different
entanglement measures.  The first of these is the entanglement or von
Neumann entropy.  The other is the linear entropy.  While the former
is more common, the latter is easier to compute and scales
monotonically with the entanglement entropy.  Figure
\ref{fig:Sandtrrho2} shows a comparison between these two measures for
$\alpha=0.2$.

The density matrix of the above described system is 
\begin{eqnarray}
	\label{eq:infldensity}
	\rho &=& \ket{\psi}\bra{\psi}  \\
	&=&\!\!\!\!\!\!\!\!
	\sum_{\mpm,\npm,m'^\pm, n'^\pm=0}^\infty\!\!\!\!\!\!\!\!\!\!\!\!\!\!\!
	B_{\mpm,\npm} B_{m'^\pm,n'^\pm}^\dagger
	\ket{\mpm}\ket{\npm}\bra{n'^\pm}\bra{m'^\pm}~~~~~~~
\end{eqnarray}
and we assume that the modes with momentum $2k$ are inaccessible to
measurement.  This gives rise to a reduced density matrix obtained
from tracing over the unobserved degrees of freedom.
\begin{eqnarray}
	\label{eq:inflreddensity}
	\rho_N &=& {\rm Tr}_M \rho = \sum_{m''^\pm =0}^\infty
	\braket{m''^\pm}{\psi}\braket{\psi}{m''^\pm} \\
	\label{eq:inflreddensity2}
	&=& \!\!\!\! \sum_{n'^\pm =0}^\infty \sum_{n^\pm=0}^\infty \left( \sum_{\mpm=0}^\infty
	B_{\mpm,\npm}B_{\mpm,n'^\pm}^\dagger\!\!\right) \ket{\npm}\bra{n'^\pm}~~~~~
\end{eqnarray}

The von Neumann entropy is then a measure of the entanglement between
the $N$ system and the unobserved $M$ system.
\begin{equation}
	\label{eq:vNentropy}
	S=-{\rm Tr}(\rho_N \ln\rho_N)=-\sum_{i=1}^{N}\rho_i \ln \rho_i
\end{equation}
where the $\rho_i$'s are the eigenvalues of the reduced density
matrix, $\rho_N$.  A system with a finite Hilbert space spanned by $N$
basis states will have a maximum entropy $S_{\rm max} = \ln{N}$.

The linear entropy, $S_L=\trpsq$, is often used as a stand-in for the
entanglement entropy since it can be computed more easily and in our
case contains the same qualitative information,
\begin{equation}
{\rm Tr}(\rho^2) = \sum_{i=1}^\infty \sum_{j=1}^i 
\begin{cases}
 2|\rho_{i j}|^2 & \text{if $j\ne i$,}
\\
|\rho_{i j}|^2 &\text{if $j=i$.}
\end{cases}
\end{equation} 
A system with a finite Hilbert space spanned by $N$ basis states will 
have a maximum linear entropy $S_{L,{\rm max}} = (N-1)/N$.  

From figure \ref{fig:Sandtrrho2}, we can see that this quantity is
nearly proportional to the entropy.  We will present the results both
in terms of entanglement entropy and $S_L$.

\subsection{Thermal Entropy and Classicality}

The amount of entropy generated can be compared to the entropy of a
thermal system that contains the same average number of particles.
For a thermal system, the entropy is
\begin{equation}
S_{\rm th} = -\sum_{n=1}^\infty \rho_{n,{\rm th}} \ln \rho_{n,{\rm th}}
\end{equation}
where the thermal density matrix is given by

\begin{equation}
\rho_{n,{\rm th}} = \frac{e^{-\beta E_n}}{\sum_{n'=1}^\infty e^{-\beta E_{n'}}}
\end{equation}
and $n'$ labels the Fock states.  Since the energy is $m=n^++n^-$,
each $n'$ state is $m+1$ times degenerate, the partition function can
be written

\begin{equation}
\sum_{n'=1}^\infty e^{-\beta E_{n'}} = \sum_{m=0}^\infty (m+1)e^{-\beta m} = \frac{1}{(e^{-\beta}-1)^2}.
\end{equation}  

\begin{figure}[h]
	\centering
		\includegraphics[width=8.5cm]{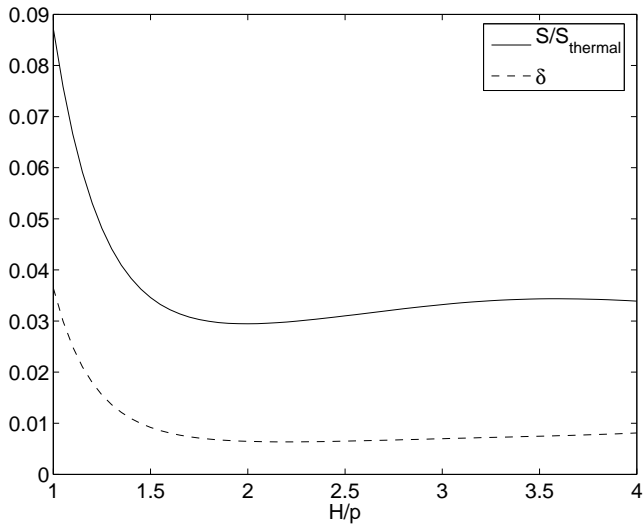}
                \caption{The fraction of information lost due to
                  tracing out the unobserved degrees of freedom,
                  defined by equation (\ref{eqn:Ilost}), and the
                  separability parameter $\delta$ (Eq.~\ref{eqn:delta2})
                  as the observed mode stretches past the horizon for
                  $\alpha = 0.2$.  By the end of the simulation,
                  $I_{\rm lost}$ appears to have leveled off to a
                  constant few percent.  For a separable (classical)
                  system, we would expect $I_{lost}$ to grow to at
                  least 0.5 to the right of the graph.  If the system
                  were to become classical, we would expect $\delta$
                  to grow to $1$.  However, it too levels off to less
                  than a percent.}
	\label{fig:Ilost}
\end{figure}

Using the relation
\begin{equation}
\label{eq:delta}
\langle n \rangle = \sum_{n'=0}^\infty n \rho_{n',\rm{th}} = \sum_{n=0}^{\infty} n(n+1)e^{-\beta n}(1-e^{-\beta})^2
\end{equation}
we can eliminate $\beta$ for $\langle n\rangle$ using
\begin{equation}
e^{-\beta} = \frac{\langle n\rangle}{2+\langle n\rangle}
\end{equation}
where $\langle n\rangle$ is the average number of particles in the
reduced system.  Finally, we can write the thermal entropy as
\begin{equation}
S_{\rm th}(\langle n \rangle) = -\sum_{m=0}^\infty (m+1)\frac{4\langle n\rangle^m}{(2+\langle n\rangle)^{m+2}}
\ln \left(\frac{4\langle n\rangle^m}{(2+\langle n\rangle)^{m+2}}\right)
\end{equation}
This quantity allows us to compare the entropy generated due to the
coupling with the total energy of a thermal system at the same
temperature.  For example, if the information content of a system is
defined as $I=S_{\rm th}-S$ then the relative information lost from
the system due to the non-linear coupling term is
\begin{equation}
	\label{eqn:Ilost}
I_{\rm lost} = 1-\frac{I}{I_{\rm max}}=\frac{S}{S_{\rm th}}.
\end{equation}  
Figure \ref{fig:Ilost} shows that the rate of information loss due to
the coupling is roughly the same as the rate of particle production.

Campo and Parentani argue that for Gaussian states at the threshold of
separability and for $\langle n \rangle \gg 1$, the entanglement
entropy between modes will be one half the entropy of the thermal
state~\cite{campo:065044}.  Since our states are not Gaussian, there
is no known general separability condition.  However, from the lack of
growth in the information loss function shown in Figure
\ref{fig:Ilost}, we can see that the Gaussian separability condition
is unlikely to occur as $\langle n \rangle$ grows much larger than $1$
at times greater than can be shown on the figure.  Therefore, these
types of non-linear interactions alone are likely insufficient to
cause the system to appear classical.

Another measure of separability used by Campo and Parentani is the
parameter $\delta$ defined by the equation
\begin{equation}
	\label{eqn:delta2}
   |c|^2 = n(n+1-\delta)
\end{equation}
where $n={\rm Tr}(\rho \create{k} \uncreate{k})=\langle n^+ \rangle$
and $c={\rm Tr}(\rho \uncreate{k} \uncreate{-k})$.  The parameter,
$\delta$, is a measure of the correlations between the $\vec{k}$ and
$-\vec{k}$ modes.  For Gaussian density matrices, it can be shown that
separability occurs when $\delta=1$.  The value of $\delta$ measured
for our model is shown alongside the information loss function in
Figure \ref{fig:Ilost}.  In both cases, the measures flatten out after
the modes leave the horizon and fail to grow as one would need for
non-linearities to explain the classicality of the quantum state.  We
can generalize the definition of $c$ to measure the correlation
between modes of different magnitudes of momenta in our system
\begin{equation}
d \equiv {\rm Tr}(\rho \uncreate{-2k} \uncreate{k} \uncreate{k}).
\label{eqn:d-def}
\end{equation}
Although the interpretation of this quantity or $\delta$ is not as
clear cut as for Gaussian density matrices, we find that both are
useful and convenient tracers of the entanglement entropy.

\subsection{Estimating the sizes of $\lambda$ and $x_{\rm final}$}
\label{sec:estim-sizes-lambda}

In order to match our above analysis with reality, we would like to
make order of magnitude estimates for the parameters $\alpha$ in
equation (\ref{eq:inflHam}) and the final value of the $x$ at the end
of inflation, $x_{\rm final}$.

For fluctuations in a scalar field other than the inflaton, the value
of $\lambda$ is essentially arbitrary; however, the gravitational
self-interaction of the field provides a strict lower bound.  Burgess,
et al. \cite{Burgess:2006} give an estimate of this self-interaction,
\begin{eqnarray}
\lambda_g &\approx& \frac{48}{(2
  \epsilon)^{3/2}}\left(\frac{H}{\PlMass}\right)^2  
= \frac{128\pi}{(2
  \epsilon)^{3/2}}  \left(\frac{M}{\PlMass}\right)^4  
\\
&\approx& 6 \times 10^{-16} \left ( \frac{\epsilon}{0.01} \right
)^{-3/2} \left ( \frac{M}{10^{14} {\rm GeV}} \right )^4
\end{eqnarray}
where $M^4$ is the vacuum energy associated with the scalar field, and 
$\epsilon = \frac{\PlMass^2}{2}\left(\frac{V'}{V}\right)^2$ is a slow-roll parameter 
which may be larger than $1$ if the scalar field is not the inflaton.  We 
have included a possible matter-dominated period following
the end of inflaton from scale factor $a_{\rm EI}$ to $a_{\rm RH}$
before reheating and taken $a$ to be the value of scale factor at the
end of inflation.    

The parameter $\alpha$ was introduced in equation (\ref{eq:inflHam}) to 
replace 
\begin{equation}
   \alpha = \frac{\lambda V H}{8 \sqrt{2\pi k \PlMass}}
\end{equation}
So, if we take, for example, a mode of size $\omega = c k \sim 0.1 {\rm Hz}=
5\times10^{-45}\PlMass$ today, we arrive at an estimate for $\alpha$ due to 
gravitational self-interactions. 
\begin{equation}
	\alpha_g \approx 2 \times 10^{-3} \left ( \frac{\epsilon}{0.01}
        \right )^{-3/2} \left ( \frac{M}{10^{14} {\rm GeV}} \right )^6
        \left ( \frac{\omega}{0.1~{\rm Hz}} \right )^{-1/2}
\end{equation}  
If the scalar field in question is the inflaton field, the
gravitational self-interaction will dominate over self-coupling
interactions.

The analysis here has assumed that reheating is quick and efficient
\cite{1997PhRvD..56.3258K,2004PhDT.......414Z}, but in principle the
end of inflaton may be followed by a period of matter domination from
scale factor $a_{\rm EI}$ to $a_{\rm RH}$ before reheating.  With this
generalization, the comoving Hubble rate at the end of inflation is
\begin{eqnarray}
a_{\rm EI} H &=&  \left ( \frac{\pi^2}{30} g_r \frac{a_{\rm EI}}{a_{\rm
      RH}}  \right )^{1/4} \left (\frac{8\pi}{3}\right )^{1/2}
\frac{T_0 M}{\PlMass} \\
&=& 6.3 \left (  g_r \frac{a_{\rm EI}}{a_{\rm
      RH}}  \right )^{1/4} \frac{M}{10^{14} {\rm GeV}}   {\rm MHz}
\end{eqnarray}
where $M^4$ is the vacuum energy associated with the inflaton field,
($\approx \lambda \PlMass^4/4$) and $g_r$ is the number of
relativistic degrees of freedom at the end of reheating where the
photon counts as two.  The value of
$x_{\rm final}$ (at the end of inflation) for the comoving scale
$a_{\rm EI} H$ is simply unity and for other scales we have
\begin{equation}
x_\rmscr{final}=\frac{a_{\rm EI} H}{\omega} = 6.3\times 10^7 g_r^{1/4}  \frac{M}{10^{14} {\rm GeV}}
\frac{0.1~\rmmat{Hz}}{\omega}
\label{eq:21}
\end{equation}
Consequently although the correlations are present on all scales, they
are most obvious on the comoving scale of the Hubble length at the end
of inflation (\ie really small scales).  On these small scales the
density fluctuations are well into the non-linear regime today but
tensor fluctuations, gravitational waves (GW), would still be a loyal
tracer of these correlations.  Inflationary tensor perturbations were 
first calculated in \cite{Starobinsky:1979}.

The expression given in Eq.~\ref{eq:21} is very uncertain.  Typically
today's Hubble scale is assumed to pass out through the Hubble length
during inflation about 50$-$60 $e-$foldings \cite{Liddle:2000};
Eq.~\ref{eq:21} gives 56 $e-$foldings before the end, so the
centihertz scale would pass through the Hubble length 12$-$22
$e-$foldings before the end of inflaton.  However, the former number
is highly uncertain.  For example, if inflation occurs at a lower
energy scale or if there is a epoch of late ``thermal inflaton''
\cite{1996PhRvD..53.1784L,1995PhRvL..75..398D,1996NuPhB.458..291D},
the number of $e-$foldings for today's Hubble scale could be as low as
25 \cite{Liddle:2000}, yielding $x_\rmscr{final} \ll\ 1$ for $ck \sim
0.1$~Hz.

Because the simulation increases in complexity as particles are
produced (see figure \ref{fig:Fockvsx}), we are confined to keeping
$x_{\rm final}\sim O(1)$.  So, even though $\alpha$ may be small in
reality, there may be sufficient time during inflation for even a
small non-linearity to produce a great deal of entanglement entropy
because of very large values of $x_{final}$.

\section{Results}

\begin{figure}[h]
	\centering
		\includegraphics[width=8.5cm]{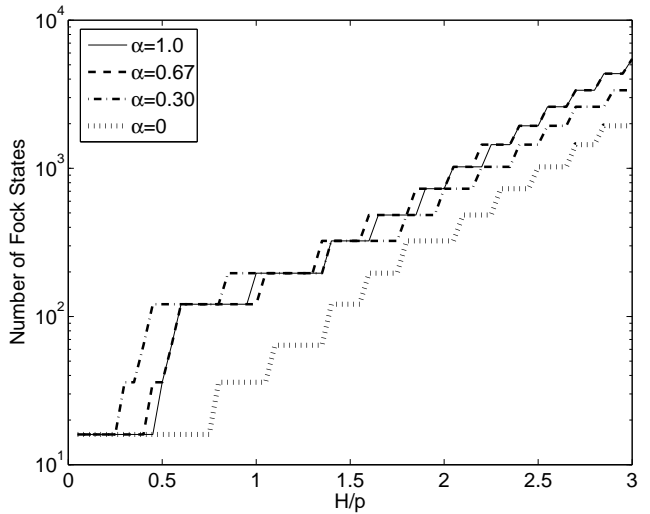}
	\caption{The number of Fock states associated with the reduced system.}
	\label{fig:Fockvsx}
\end{figure}

We would like to investigate how the amount of entropy generated in a
single mode scales with the coupling strength and the duration of
inflation (i.e.\ $\alpha$ and $x_{\rm final}$).  Figure
\ref{fig:Sandtrrho2} explicitly shows the creation of entanglement
entropy for $\alpha=1$ as the universe undergoes its inflationary
phase.  The horizontal axis, $x=H/p$, is the physical size of a mode
with respect to the horizon scale.  The entanglement entropy increases
less quickly than exponentially, which would be a straight line on the
figure.  Unfortunately, as was mentioned previously, the computational
size of the problem prevents us from simulating far past horizon
crossing because the number of particles becomes too large.  Figure
\ref{fig:Fockvsx} shows how many Fock states are in the reduced system
at each time step in the simulation.  The number of states being
integrated is this number to the $3/2$ power, and the number of
entries in the density matrix is the square of this number.

\begin{figure}[h]
	\centering
        \includegraphics[width=8.5cm]{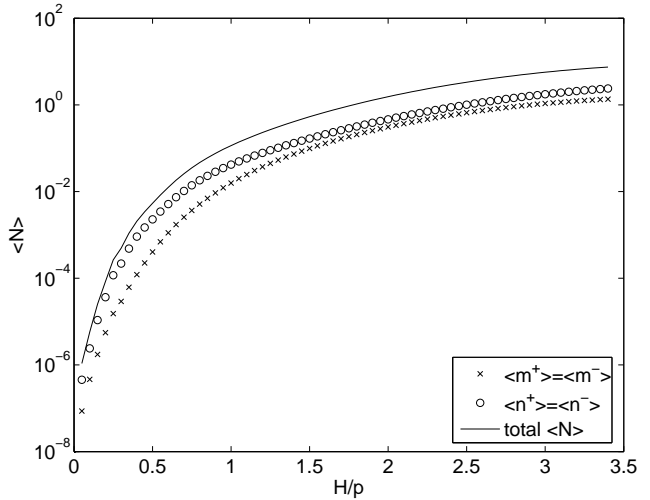}
	\caption{Evolution of the average particle numbers for each
          mode for $\alpha=1.0$}
	\label{fig:Nvsx}
\end{figure}

\begin{figure}[h]
	\centering
		\includegraphics[width=8.5cm]{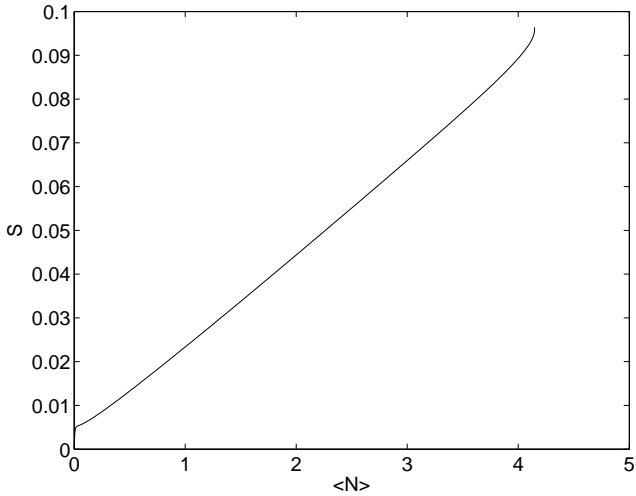}
                \caption{Entanglement entropy vs. average particle
                  number for $\alpha=0.2$}
	\label{fig:SvsN}
\end{figure}

The evolution of particles in the system is shown in figure
\ref{fig:Nvsx}.  Our results are consistent with those found in
Heyl~\cite{Heyl:2006fb} and show a nearly exponential evolution of the
average particle number.  Moreover, we can look at the evolution of
each mode separately.  For $\lambda=0$, each mode evolves
according to the same equations of motion, and in this case, there is
no difference between the rate that each of the modes evolves.
However, the nature of the interaction between the modes is not
symmetric because the decay of a single $M$ mode particle results in 2
$N$ mode particles and therefore the interaction results in an
increased rate of production of $N$ mode particles, relative to the
$M$ mode.  Figure \ref{fig:SvsN} shows how the entanglement entropy 
scales with average particle number when $\alpha=0.2$.  

\begin{figure}[h]
	\centering
		\includegraphics[width=8.5cm]{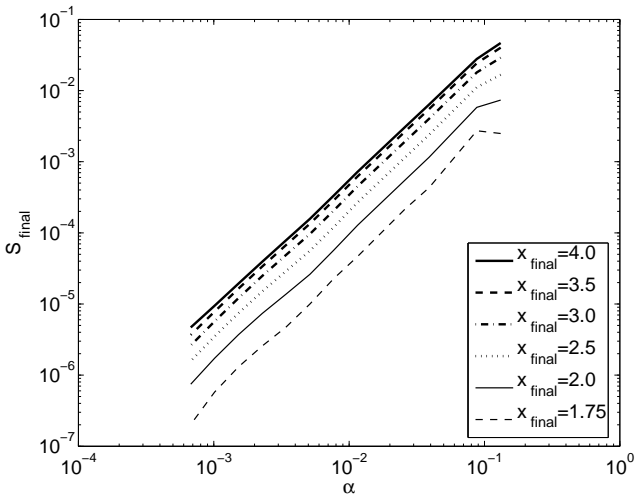}
	\caption{von Neumann entropy vs. $\alpha$ for various values of $x_{\rm final}$}
	\label{fig:bigalphaplot}
\end{figure}

We performed the simulation for a variety of values for the coupling,
$\alpha$, spanning several orders of magnitude.  Figure
\ref{fig:bigalphaplot} shows entropy generation as a function of
$\alpha$ for a variety of inflation durations $x_{\rm final}$.  From
this plot, we can see that $S_{\rm final}$ scales roughly as a power
law in $\alpha$.  Most of the $\alpha$ dependence can be removed by
dividing $S_{\rm final}$ by $\alpha^{1.75}$.  Doing this also helps to
illustrate how $S_{\rm final}$ scales with $x_{\rm final}$.  As
expected, there is no entropy generated without the coupling terms
(i.e.\ when $\alpha=0$).  In this case, there is no communication
between modes of the scalar field and they evolve independently.

\begin{figure}[h]
	\centering
		\includegraphics[width=8.5cm]{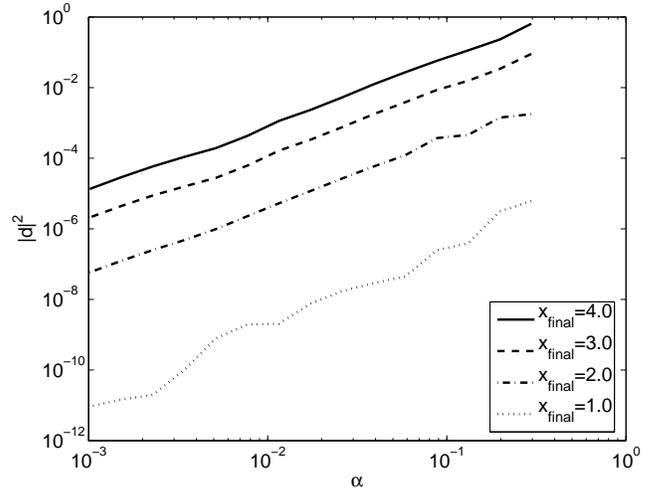}
                \caption{$|d|^2 = |{\rm Tr}(\rho
                  \uncreate{2k}\uncreate{k}\uncreate{k})|^2$ scales
                  with $\alpha$ in much the same way as $S$, but is
                  less costly to compute.}
	\label{fig:CPdvsl}
\end{figure}
\begin{figure}[h]
	\centering
		\includegraphics[width=8.5cm]{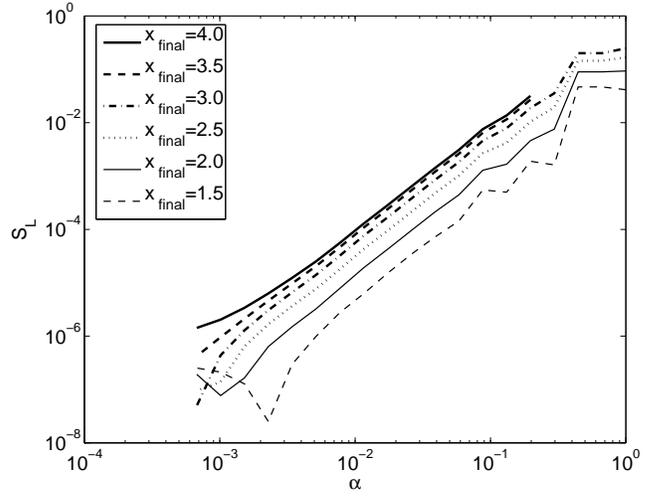}
	\caption{$S_L$ vs. $\alpha$ for various values of $x_{\rm final}$}
	\label{fig:bigalphaplotp2}
\end{figure}

As was mentioned earlier, $S_L$ is a useful stand-in for $S$ that can
be computed faster than $S$.  Figure \ref{fig:bigalphaplotp2} echoes
the previous results in terms of $S_L$ instead of $S$.  In this case,
$\trpsq$ scales more like $\alpha^2$ instead of $\alpha^{1.75}$.
However, both $S_L$ and $S$ demonstrate the same qualitative
behaviour.

\begin{figure}[h]
	\centering
		\includegraphics[width=8.5cm]{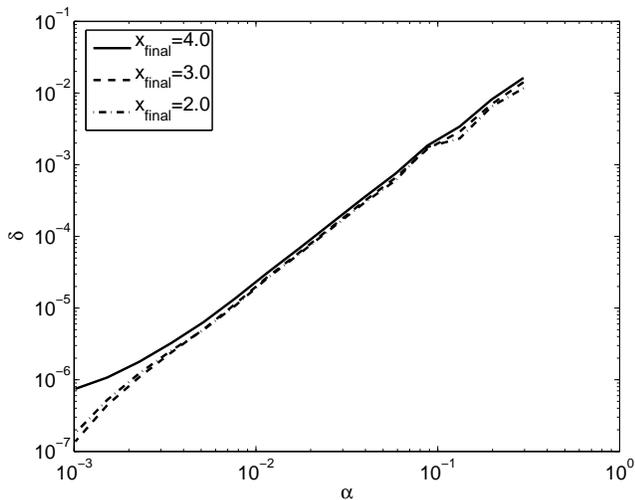}
                \caption{$\delta$, defined in equation
                  (\ref{eqn:delta2}) scales with $\alpha$ in much the
                  same way as $S_L$, but is less costly to compute.}
	\label{fig:CPdeltavsl}
\end{figure}

In addition, we have found other useful stand-ins for the entanglement
entropy that are easier to compute and scale similarly with $\alpha$.
The $\delta$ parameter, defined in (\ref{eqn:delta2}), scales roughly
like an $\alpha^2$ power law much like $S_L$.  Figure
\ref{fig:CPdeltavsl} shows the power law behaviour of this function.
Additionally, if we use a simple measure of correlation between
entangled modes, $|d|^2$ (Eq.~\ref{eqn:d-def}), we find that its
scales like $\alpha^{1.85}$ (see fig. \ref{fig:CPdvsl}), and so can be
a useful stand-in for the von Neumann entropy, $S \propto
\alpha^{1.75}$.

In the real universe, we are dealing with small values of $\lambda$
and very large values of $x$.  However, the simulation outlined in
this paper is limited because its computational complexity increases
dramatically as particles are produced, even for small values of the
coupling, $\alpha$.  Moreover, for small values of $\alpha$, the
production of entropy is too small to be meaningful.  While the
dependence of $S$ on $\alpha$ nearly follows a power law, there is no
simple relation describing the dependence of $S$ on $x_{\rm final}$.
The value $S_L$ is approximately proportional to $\alpha^2 x^3$
over a wide range of $\alpha$ and the modest range $x$ probed by the
simulations;
therefore, very roughly, we can write the scaling law as
$S_L \propto \alpha^{2} x_{\rm final}^3$ where
\begin{equation}
S_L \approx 10^{16} g_r^{3/4} \left ( \frac{M}{10^{14} {\rm
      GeV}} \right )^{15} \left ( \frac{\omega}{0.1~{\rm Hz}} \right
  )^{-4} .
\end{equation}
Of course, only values of $S_L$ less than unity make sense, so a
larger value from the fitting formula indicates that $S_L$ is very
close to one.  However, a value of $S_L < 1$ is obtained
by lowering the mass scale of inflation below
\begin{equation}
M < 8\times 10^{12} \left ( \frac{\omega}{0.1~{\rm Hz}} \right
)^{-4/15} 
{\rm GeV};
\end{equation}
therefore, if the energy scale of inflation is low, the quantum states
of fluctuations at $\omega \sim 0.1$~Hz will remain coherent despite
the non-linear coupling.


The simulation was checked for consistency in several ways.  First, we
traced the probability throughout the simulation measured both by the
sum of squares of the matrix elements $\sum_{\mpm,\npm = 0}^\infty
A_{\mpm,\npm}$ and the trace of the density operator.  
Both of these quantities were conserved
to a few parts in $10^{-7}$.  Moreover, we estimated the level of
numerical error by rerunning the simulation with a variety of phase
rotations multiplying the initial wavefunction.  The standard
deviation of the results from these numerical changes in the initial
conditions give us an idea of the level of numerical error in the
simulation, which were typically at the level of one part per
thousand.


\section{Conclusions}

In this paper we have developed a model in which two modes of a scalar
field evolve during inflation and we have computed the entanglement
entropy between them.  The entanglement entropy generated between
observed and unobserved modes in the inflaton field give the
appearance that entropy is being produced, even though the scalar
field remains in an overall pure state.  The preceding results clearly
show that non-linearities in the inflaton potential give rise to a
generation of entanglement entropy between observed modes and
unobserved modes in a scalar field during inflation.  This entropy is
an additional source to that caused by coupling to external degrees of
freedom \cite{Kiefer:2006je}, entanglement between the inside and
outside of the horizon \cite{Sharman2007} and that which is created
during reheating after inflation has ended.

We have attempted to extrapolate the results of our simulation to the
real universe.  The relevant parameters determining the amount of
entropy generated via non-linearities are the strength of the coupling
$\alpha_g \sim 10^{-3}$ and the scale of the fluctuation at the end of
inflation given by the dimensionless parameter $x_{\rm final} \sim
10^{7}$.  The entanglement entropy was found to scale like
$\alpha^{1.75}$ for a fixed $x_{\rm final}$.  The dependence of $S_L$
on $x_{\rm final}$ for a given value of $\alpha$ is not as
straightforward, but $S_L \propto x_{\rm final}^3$ over a short range
of $x_{\rm final}$ values.  Based on these rough scaling patterns, we
estimate that non-linearities due to gravity and inflaton
self-coupling are insufficient to decohere modes that spend only a
few Hubble times at super-horizon scales.  In particular, if the
energy scale of inflaton is less than $10^{13}$~GeV, fluctuations 
at about 0.1~Hz may remain coherent.

We found two measures of the decoherence related to the correlations
between modes of different momenta provide a faithful estimate of the
entanglement entropy in our model --- one of these measures is new to
this work ($d$) and specifically probes the non-linear coupling
between modes.  In particular these estimates are very inexpensive to
calculate as compared to the von Neumann entropy and should prove
useful for more detailed models of entropy generation.

It is usually assumed that the main contribution to the entropy
observed in the density perturbations is generated during reheating,
when the inflaton decays.  However, the analysis demonstrates that
entropy can be generated independently of reheating provided there is
even a small non-linearity in the scalar potential; therefore, the
results are applicable to scalar fields that do not participate in
reheating.  For example, the gravitational wave background can be
treated as a pair of scalar fields, so even tensor fluctuations may
contribute to the entropy and the classicality of the distribution of
density perturbations in this way and observations of the
gravitational wave background at high frequency could reveal the
quantum mechanical origin of density fluctuations.

\bigskip
\paragraph*{Acknowledgments}

This research was supported by funding from NSERC.  The calculations
were performed on computing infrastructure purchased with funds from
the Canadian Foundation for Innovation and the British Columbia
Knowledge Development Fund.



\bibliographystyle{unsrt}
\bibliography{references}


\end{document}